\documentclass[aip,jap,reprint]{revtex4-1}%
\usepackage{graphicx}
\usepackage{dcolumn}
\usepackage{bm}
\usepackage{amsmath}%
\usepackage{amsfonts}%
\usepackage{amssymb}

\begin{document}

\title
{Photon-induced carrier transport in high efficiency mid-infrared quantum cascade lasers}
\author{Alp\'ar M\'aty\'as}
\email{alparmat@mytum.de}
\homepage{http://www.nano.ei.tum.de/noether}
\affiliation
{Emmy Noether Research Group ``Modeling of Quantum Cascade Devices``, Technische Universit\"{a}%
t M\"{u}nchen, D-80333 Munich, Germany}
\affiliation{Institute for Nanoelectronics, Technische Universit\"{a}%
t M\"{u}nchen, D-80333 Munich, Germany}
\author{Paolo Lugli}
\affiliation{Institute for Nanoelectronics, Technische Universit\"{a}%
t M\"{u}nchen, D-80333 Munich, Germany}
\author{Christian Jirauschek}
\affiliation
{Emmy Noether Research Group ``Modeling of Quantum Cascade Devices``, Technische Universit\"{a}%
t M\"{u}nchen, D-80333 Munich, Germany}
\affiliation{Institute for Nanoelectronics, Technische Universit\"{a}%
t M\"{u}nchen, D-80333 Munich, Germany}
\date{\today, published as J. Appl. Phys. 110, 013108 (2011)}
\begin{abstract}
A mid-infrared quantum cascade laser with high wall-plug efficiency is analyzed by means of an ensemble Monte-Carlo method. Both the carrier transport and the cavity field dynamics are included in the simulation, offering a self-consistent approach for analyzing and optimizing the laser operation. It is shown that at low temperatures, photon emission and absorption can govern the carrier transport in such devices. Furthermore we find that photon-induced scattering can strongly affect the kinetic electron distributions within the subbands. Our results are validated against available experimental data.
\end{abstract}
\maketitle

\section{INTRODUCTION}

Quantum cascade lasers (QCLs) are among the most promising mid-infrared (MIR)
laser sources, offering applications in gas sensing, free space communications
and spectroscopy. Since the first operating device was presented in
1994,\cite{1994Sci...264..553F} QCL designs have been constantly improved with
respect to their efficiency and output power. Recently QCLs with wall-plug
efficiencies (WPEs) of around 50\% were reported for the first
time.\cite{2010NaPho...4...95L,2010NaPho...4...99B} In such structures, light
emission and absorption are not only relevant with respect to the generated
optical power, but also strongly affect the carrier transport in the devices.
In fact, for the low temperatures where these high efficiencies are reached,
the photon-induced processes dominate the other scattering mechanisms. Thus,
to adequately model the operation of these lasers, the optical cavity field
has to be considered in the simulation. While this is routinely done in
one-dimensional
simulations,\cite{2010OExpr..1813616G,2010JAP...108j3108S,2010NJPh...12c3045T,2011JAP...109a3111B}
the cavity field is usually neglected in fully three-dimensional (3-D)
approaches like the ensemble Monte-Carlo (EMC),~\cite{2001ApPhL..78.2902I}
non-equilibrium Green's functions
(NEGF)\cite{2005ApPhL..86d1108B,2009PhRvB..79s5323K} or 3-D density
matrix\cite{2006ApPhL..89i1112W,PhysRevLett.87.146603} method. However, such
3-D simulations, where the in-plane carrier dynamics is explicitly considered,
do not only yield level occupations, but also the kinetic carrier
distributions within these levels. Here, we employ the EMC method, which has
has been intensely used to investigate the carrier transport in both
MIR\cite{2001ApPhL..78.2902I,2002ApPhL..80..920C,2007JAP...101f3101G,2007JAP...102k3107G,2008JAP...103g3101G,2010JAP...108g3106B}
and
terahertz\cite{2001ApPhL..79.3920K,2003ApPhL..83..207C,2005JAP....97d3702B,2007JAP...101h6109J,2008JAP...104d3101L,2010ApPhL..96t1110M}
QCLs. To include also the optical processes, we have recently extended this
approach, allowing for self-consistent coupled simulations of the carrier
transport and the optical cavity field.\cite{2010ApPhL..96a1103J} The EMC
method is a semiclassical approach, i.e., quantum correlations are neglected
in contrast to NEGF\cite{Matyas2009} or density
matrix\cite{PhysRevLett.87.146603} calculations; however, the carrier
transport in MIR QCLs has been shown to be largely incoherent.\cite{PhysRevLett.87.146603}

The goal of the present study is to analyze the carrier transport and lasing
operation in a record-efficiency MIR QCL,\cite{2010NaPho...4...99B} with a
particular focus on the influence of photon-induced scattering on the carrier
transport. Specifically, we show that the inclusion of light emission and
absorption in the simulation is crucial to obtain a realistic description for
such devices. Furthermore, our analysis provides insight into the carrier
dynamics on a microscopic level, for example the kinetic electron
distributions in the upper and lower laser level which are hardly accessible
to experimental observation. The paper is organized as follows: In Section
\ref{method}, we give a basic overview of our Monte-Carlo approach,
specifically adapted for InGaAs/InAlAs strain-compensated MIR QCLs. In Section
\ref{results}, we compare our simulation results to available experimental
data and demonstrate the strong influence of stimulated emission on the
carrier transport. The paper is concluded in Section \ref{conclusions}.

\section{\label{method}METHOD}

The EMC method is based on the semiclassical Boltzmann transport
equation.\cite{PhysRevLett.87.146603} Scattering is self-consistently
accounted for based on Fermi's golden rule. All the relevant mechanisms like
electron (e)-longitudinal optical (LO) and acoustic phonon, e-interface
roughness, e-impurity and e-e scattering are routinely considered in our
simulation tool.\cite{2009JAP...105l3102J,2010JAP...107a3104J} Moreover,
various effects relevant for MIR QCLs based on the\ InGaAs/InAlAs material
system have been added. We have included random alloy
scattering,\cite{2003JAP....93.1586U} with a scattering potential of
$0.3\,\mathrm{eV}$ reported for high indium content
InGaAs.\cite{1998JAP....84.2112R} Furthermore, we account for InAs- and
GaAs-like phonons, using their composition dependent values for the phonon
energy.\cite{1999JAP....85.7276A} The scattering rates are weighted by the
concentration of the individual materials (InAs and GaAs). The influence of
the AlAs-like branch is believed to be negligible in QCL
structures.\cite{2005ApPhL..87g2104D} Here, the bulk phonon approximation is
adopted, which was shown to be a valid approach for the simulation of such QCL\ structures.\cite{2008JAP...103g3101G}

The (parallel and perpendicular) effective masses have been implemented
considering strain\cite{1993PhRvB..48.8102S} and non-parabolicity. Our
implementation of non-parabolicity is based on the approach developed by
Ekenberg.\cite{PhysRevB.40.7714} Non-parabolicity parameters were determined
from the material bandgap,\cite{1987PhRvB..35.7770N} using temperature
dependent values.\cite{2001JAP....89.5815V} In the InGaAs material system, the
parallel non-parabolicity is enhanced by a factor of $1.7$ as compared to the
perpendicular value.\cite{1993PhRvB..48.2328H} The perpendicular effective
mass affects the subband energies and wavefunctions, as considered in our
Schr\"{o}dinger-Poisson solver.\cite{2009IJQE...45..1059J} The parallel
effective mass is accounted for by assigning a different value to each
subband, affecting the scattering rates in the EMC solver. Here we focus on
simulations at a lattice temperature of $40\,\mathrm{K}$ where the
investigated structure operates with a record wall-plug efficiency of above
50\%.\cite{2010NaPho...4...99B} At such low temperatures, the kinetic electron
energies are still moderate, whereas for room temperature operation, a more
complex implementation of nonparabolicity might be required, e.g., based on
\textbf{k%
$\cdot$%
p} theory.\cite{2007JAP...101f3101G,2007JAP...102k3107G,2008JAP...103g3101G}
Furthermore, at low temperatures, the electron leakage into indirect valleys,
not considered in our simulations, is very small.\cite{2007JAP...101f3101G,2007JAP...102k3107G}

The interface roughness is typically described by a mean height $\Delta$ and a
correlation length $\Lambda$. In contrast to the well-known bulk material
parameters, this quantity is hardly accessible to experimental measurement and
depends critically on the growth conditions. Thus, there is an uncertainty
regarding the values of $\Delta$ and $\Lambda$%
.\cite{2008PSSCR...5..232K,2009JAP...105l3102J} However, experimental data
indicate that $\Delta\Lambda\approx1\,\mathrm{nm}^{2}$ for the InGaAs/InAlAs
structures,\cite{2005ApPhL..86f2113T,2008ApPhL..93n1103W} reducing the
uncertainty to a single parameter value. We choose $\Delta=0.06\,\mathrm{nm}$,
which yields the best agreement with the experimental results. This value is
somewhat lower than previously used values for strain-free lattice-matched
structures.\cite{2005ApPhL..86f2113T,2006ApPhL..89q2120V} However, we note
that vertical correlations, which are not included in our simulations, can
reduce the effect of interface roughness for strained (e.g., strain-balanced)
quantum cascade lasers,\cite{2005ApPhL..86f2113T} as considered here.

Lasing is implemented based on a recently published approach, treating the
photon dynamics in terms of classical intensity evolution equations and
accounting for photon-induced scattering in the EMC
solver.\cite{2010ApPhL..96a1103J,oexprJirau01} In this way we can
self-consistently describe the coupled carrier-light dynamics due to
absorption as well as stimulated and spontaneous emission. For the
investigated design operating at $5\,\mu\mathrm{m}$,\cite{2010NaPho...4...99B}
the mirror loss, which amounts to $6.4\,\mathrm{cm}^{-1}$ for a
$2\,\mathrm{mm}$ long structure, dominates the waveguide loss, which is about
$0.5\,\mathrm{cm}^{-1}$ for such cavities.\cite{5069088} The confinement
factor is chosen to be $0.8$ as found for a similar design.\cite{5069088} For
our simulation, we use 1200 longitudinal modes in the frequency range between
$50$ and $80\,\mathrm{THz}$, corresponding to a Fabry-Perot mode spacing of
$25\,\mathrm{GHz}$.

\section{\label{results}RESULTS}

Results are presented for a recently fabricated high efficiency QCL operating
at $5\,\mu\mathrm{m}$.\cite{2010NaPho...4...99B} The simulations were
performed at a lattice temperature of $40\,\mathrm{K}$, where the record WPE
of 53\% was observed.

\begin{figure}[ptb]
\includegraphics{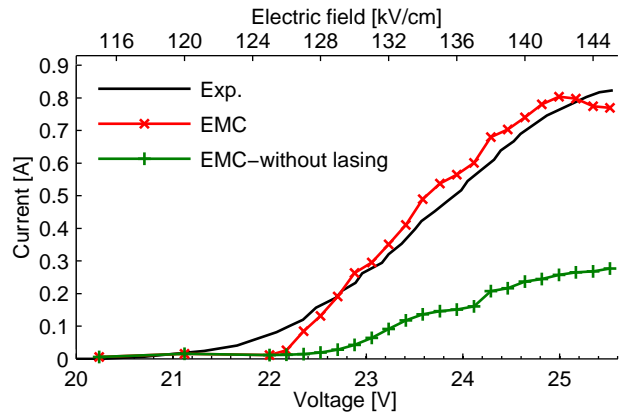}\caption{(Color online) Current-voltage
characteristics. The EMC simulation results with (X marks) and without
(crosses) lasing included are compared to available experimental
data\cite{2010NaPho...4...99B} (solid curve). The electric current is governed
by stimulated photon emission and absorption processes, as has to be expected
for a WPE as high as 50\%.}%
\label{fig1}%
\end{figure}

In Fig.~\ref{fig1}, we compare the current-voltage characteristics provided by
EMC simulations to experiment. The simulations were performed at biases
ranging from $115\,\mathrm{kV}/\mathrm{cm}$ to $145\,\mathrm{kV}/\mathrm{cm}$.
For comparison to experiment, these were converted to the voltage points in
Fig.~\ref{fig1}, by considering 80 stages with a thickness of
$22.1\,\mathrm{nm}$ each.\cite{2010NaPho...4...99B} Above threshold, good
agreement is found if lasing is included, while the current due to
non-radiative processes (EMC without lasing) is lower by a factor of almost 3
than the experimentally measured current. This shows that stimulated processes
become more and more important for a correct description of the carrier
transport as the WPE of QCLs is improved. On the other hand, the spontaneous
photon emission rates in our simulation are far too low to affect\ the carrier
transport, which is in agreement with theoretical
considerations.\cite{2008IJQE...44...12Y} The onset of the negative
differential resistance (NDR) regime agrees well, occuring at
$25.1\,\mathrm{V}$ for the EMC with lasing included and $25.6\,\mathrm{V}$ in
the experiment. For low fields where the energy levels are not aligned, the
simulation underestimates the experimentally observed current. Here, the
scattering-induced transport is not efficient, and the remaining current can
likely be attributed to coherent low-field transport which is not included in
the EMC simulation.~\cite{Matyas2009} For design optimization with respect to
the WPE, the parasitic channels should be suppressed and the stimulated
emission into the lasing modes maximized. Such a task can only be performed
with an approach taking into account the optical cavity field.

\begin{figure}[ptbptb]
\includegraphics{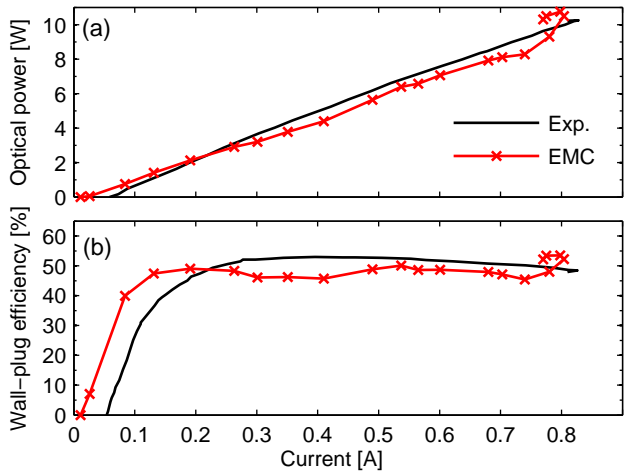}\caption{(Color online) (a) Current-optical power
and (b) current-WPE characteristics. The EMC simulation results with lasing
included (X marks) are compared to available experimental
data\cite{2010NaPho...4...99B} (solid curve).}%
\label{fig2}%
\end{figure}

In Fig.~\ref{fig2}(a) and (b) we compare the simulated and experimental
current-output power and current-WPE characteristics. In the EMC simulation,
the bias dependent WPE $\eta_{\mathrm{WPE}}$ is computed as $\eta
_{\mathrm{WPE}}=P_{\mathrm{opt}}/P_{\mathrm{el}}$. Here, $P_{\mathrm{opt}}$ is
the simulated optical power emitted through both facets as in the
experiment,\cite{2010NaPho...4...99B} and the electric power $P_{\mathrm{el}}%
$\ is the product of the applied voltage and the simulated electric current.
The simulated and experimental current-output power characteristics in
Fig.~\ref{fig2}(a) show excellent qualitative and quantitative agreement. The
maximum emitted optical power is about $10\,\mathrm{W}$, which is in both
cases obtained around the onset of NDR, where the current reaches its maximum
value of $0.8\,\mathrm{A}$. For higher biases, i.e., in the NDR regime, the
simulated optical power and electric current get reduced again. The simulated
threshold current is lower than the experimental value, for the reasons
discussed in the previous paragraph. Also the simulated and experimental
current-WPE characteristics shown in Fig.~\ref{fig2}(b) agree well.
Particularly, the maximum simulated WPE of 49\% below the onset of NDR
compares very well to the experimental value of 53\%. The simulated high WPE
value of 53.5\% around the onset of NDR is not observed in the experiment,
which we attribute to the fact that the operation in the NDR region is
unstable due to domain formation.\cite{2010ApPhL..97h1105W,2011JAP...109g3112W}

\begin{figure}[ptbptbptb]
\includegraphics{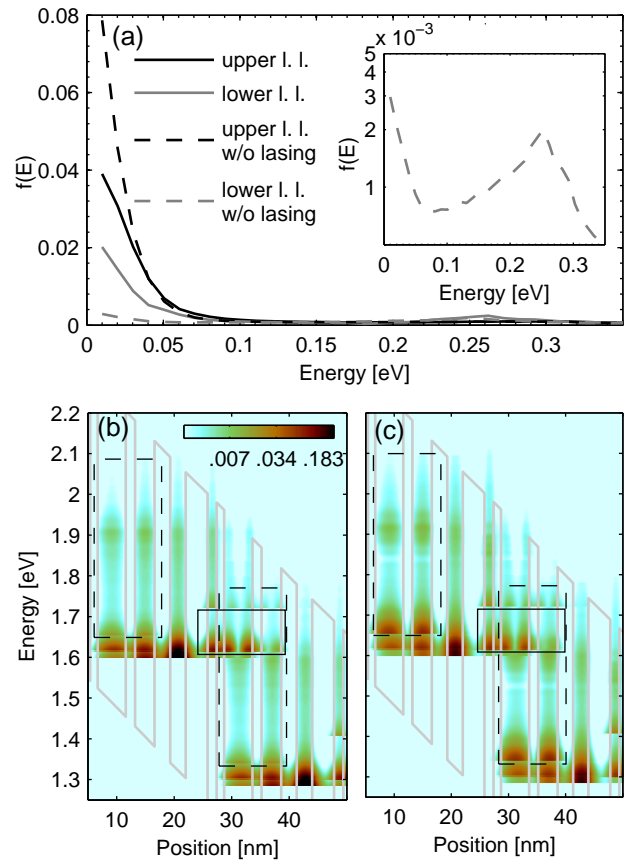}\caption{(Color online) (a) Simulated kinetic
electron distribution $f\left(  E\right)  $ in the upper and lower laser level
with and without lasing included. (b) Energy resolved electron density without
lasing included; (c) energy resolved electron density with lasing included.
The upper laser levels are marked by solid rectangles, and the lower laser
level is marked by dashed rectangles.}%
\label{fig3}%
\end{figure}

Full $k$-space three-dimensional simulation approaches like EMC can yield
information on the microscopic level, which is hardly accessible to
experimental observation. In the following, we investigate the intrasubband
kinetic carrier distributions. These can be characterized by corresponding
electron temperatures only in the case of quasi-thermal equilibrium within the
subbands, corresponding to a Maxwellian distribution for low doping. The
kinetic electron distribution in the upper and lower laser level is shown in
Fig.~\ref{fig3}(a). The bias is $25.1\,\mathrm{V}$, where the simulated
current and output power reach their maximum values. We note that for this
bias, optical transitions from two upper levels contribute significantly to
lasing. For simplicity, we restrict our discussion to one of these states,
since the kinetic electron distribution function is found to be similar for
the other level. The carrier distributions in the laser levels change
significantly by taking lasing into account (solid curves), as compared to the
case without lasing (dashed curves). The lasing action leads to a depletion of
the upper laser levels and a filling of the lower laser level, corresponding
to the effect of gain saturation. In the inset of Fig.~\ref{fig3}(a), the
electron distribution in the lower laser level without lasing is shown on a
logarithmic scale, i.e., a Maxwellian distribution would produce a straight
line. The distribution is highly non-Maxwellian with an additional peak at
around $250\,\mathrm{meV}$, corresponding to the energy spacing between upper
and lower laser level. This bump stems from nonradiative transitions from the
upper laser level, mainly LO phonon scattering as the dominant nonradiative
mechanism. The energetic extension of the bump is partly due to the kinetic
electron distribution in the upper laser level and the finite phonon energies
of $29.5\,\mathrm{meV}$ and $32.2\,\mathrm{meV}$ for the two LO\ branches
considered here. E-e scattering, which is the predominant intrasubband
scattering mechanism, is unable to thermalize the intrasubband carrier
distribution strongly enough to suppress the
bump.\cite{2001ApPhL..78.2902I,2008pssc..5..221J} The inclusion of lasing
action leads not only to a filling of the lower laser level, but also to a
more thermalized kinetic electron distribution, while the bump at around
$250\,\mathrm{meV}$ still persists. A least square fit produces an electron
temperature of $T_{\mathrm{e}}=314\,\mathrm{K}$ (upper laser level) and
$T_{\mathrm{e}}=344\,\mathrm{K}$ (lower laser level) with lasing included.
This is consistent with the observation that in strain compensated structures
the electronic temperature is clearly above the lattice temperature.\cite{2007ApPhL..91p1111V,2010NJPh...12c3045T}

In Fig.~\ref{fig3}(b) and (c), the energy resolved electron density
(normalized to its maximum value) is shown without and with lasing included,
again for a bias of $25.1\,\mathrm{V}$. For the two quantum wells located
between $6\,\mathrm{nm}$ and $18\,\mathrm{nm}$, the upper laser level is
omitted, so that the high-energy peak discussed in the previous paragraph (see
inset of Fig.~\ref{fig3}(a)), here located at around $1.9\,\mathrm{eV}$, is
clearly visible. By comparing Fig.~\ref{fig3}(b) and (c), we can observe the
changes in the energy resolved electron density for the laser levels without
and with lasing included. The high-energy tails of the kinetic electron
distributions remain basically unaffected. In particular, the additional
high-energy peak for the lower laser level appears also if lasing is accounted for.

We have successfully tested our approach for a further high efficiency QCL
design\cite{2010NaPho...4...95L} using the same material parameters, again
finding good agreement with experiment.

\section{\label{conclusions}CONCLUSION}

Based on a self-consistent EMC carrier transport simulation including the
optical cavity field, we have analyzed the effect of photon emission and
absorption on the carrier transport in a high WPE quantum cascade laser. In
the regime where efficient lasing is obtained, we find that the inclusion of
photon-induced scattering is crucial for the correct calculation of the device
current. Furthermore, a comparison to experimental data yields very good
agreement for the optical output power and WPE. An analysis of those
quantities, as also needed for design optimization, is only possible with an
approach which includes both the carrier transport and the optical cavity
field. The EMC method also enables us to investigate microscopic quantities
such as the intrasubband kinetic carrier distributions, hardly accessible to
experimental observation. The upper and lower laser level carrier
distributions are strongly affected by the lasing action and approach each
other, corresponding to gain saturation. We observe strong deviations from an
equilibrium distribution especially for the lower laser level, where a
high-energy peak in the electron distribution is found, caused by parasitic
transitions from the upper laser level. Our results show that the chosen
approach is well suited to model high efficiency MIR QCLs on a qualitative and
quantitative level, and to analyze the laser operation on a microscopic scale
which is hardly accessible to experimental observations.

\section{ACKNOWLEDGMENTS}

A. M. and C. J. acknowledge support from the Emmy Noether program of the
German Research Foundation (DFG, JI115/1-1) . We acknowledge computational
resources from the LRZ (t8481ag). A. M. additionally acknowledges support from
the TUM Graduate School.


\begin{thebibliography}{9}                                                                                                %
\bibitem {1994Sci...264..553F}J.~{Faist}, F.~{Capasso}, D.~L. {Sivco},
C.~{Sirtori}, A.~L. {Hutchinson}, and  A.~Y. {Cho}, \newblock Science
\textbf{264}, 553 (1994).

\bibitem {2010NaPho...4...95L}P.~Q. {Liu}, A.~J. {Hoffman}, M.~D. {Escarra},
K.~J. {Franz}, J.~B. {Khurgin},  Y.~{Dikmelik}, X.~{Wang}, {J.-Y.} {Fan}, and
C.~F. {Gmachl}, \newblock Nat. Photonics \textbf{4}, 95 (2010).

\bibitem {2010NaPho...4...99B}Y.~{Bai}, S.~{Slivken}, S.~{Kuboya}, S.~R.
{Darvish}, and M.~{Razeghi}, \newblock Nat. Photonics \textbf{4}, 99 (2010).

\bibitem {2010OExpr..1813616G}V.-M. {Gkortsas}, C.~{Wang}, L.~{Kuznetsova},
L.~{Diehl}, A.~{Gordon},  C.~{Jirauschek}, M.~A. {Belkin}, A.~{Belyanin},
F.~{Capasso}, and F.~X. {K{\"a}rtner}, \newblock Opt. Express \textbf{18},
13616 (2010).

\bibitem {2010JAP...108j3108S}L.~{Schrottke}, M.~{Wienold}, M.~{Giehler},
R.~{Hey}, and H.~T. {Grahn}, \newblock J. Appl. Phys. \textbf{108}, 103108 (2010).

\bibitem {2010NJPh...12c3045T}R.~{Terazzi} and J.~{Faist}, \newblock New J.
Phys. \textbf{12}, 033045 (2010).

\bibitem {2011JAP...109a3111B}G.~{Beji}, Z.~{Ikoni{\'c}}, C.~A. {Evans},
D.~{Indjin}, and P.~{Harrison}, \newblock J. Appl. Phys. \textbf{109}, 013111 (2011).

\bibitem {2001ApPhL..78.2902I}R.~C. {Iotti} and F.~{Rossi}, \newblock Appl.
Phys. Lett. \textbf{78}, 2902 (2001).

\bibitem {2005ApPhL..86d1108B}F.~{Banit}, S.-C. {Lee}, A.~{Knorr}, and
A.~{Wacker}, \newblock Appl. Phys. Lett. \textbf{86}, 041108 (2005).

\bibitem {2009PhRvB..79s5323K}T.~{Kubis}, C.~{Yeh}, P.~{Vogl}, A.~{Benz},
G.~{Fasching}, and C.~{Deutsch}, \newblock Phys. Rev. B \textbf{79}, 195323 (2009).

\bibitem {2006ApPhL..89i1112W}C.~{Weber}, F.~{Banit}, S.~{Butscher},
A.~{Knorr}, and A.~{Wacker}, \newblock Appl. Phys. Lett. \textbf{89}, 091112 (2006).

\bibitem {PhysRevLett.87.146603}R.~C. Iotti and F.~Rossi, \newblock Phys. Rev.
Lett. \textbf{87}, 146603 (2001).

\bibitem {2002ApPhL..80..920C}F.~{Compagnone}, A.~{di Carlo}, and P.~{Lugli},
\newblock Appl. Phys. Lett. \textbf{80}, 920 (2002).

\bibitem {2007JAP...101f3101G}X.~{Gao}, D.~{Botez}, and I.~{Knezevic},
\newblock J. Appl. Phys. \textbf{101}, 063101 (2007).

\bibitem {2007JAP...102k3107G}X.~{Gao}, M.~{D'Souza}, D.~{Botez}, and
I.~{Knezevic}, \newblock J. Appl. Phys. \textbf{102}, 113107 (2007).

\bibitem {2008JAP...103g3101G}X.~{Gao}, D.~{Botez}, and I.~{Knezevic},
\newblock J. Appl. Phys. \textbf{103}, 073101 (2008).

\bibitem {2010JAP...108g3106B}P.~{Borowik}, J.-L. {Thobel}, and L.~{Adamowicz}%
, \newblock J. Appl. Phys. \textbf{108}, 073106 (2010).

\bibitem {2001ApPhL..79.3920K}R.~{K{\"o}hler}, R.~C. {Iotti}, A.~{Tredicucci},
and F.~{Rossi}, \newblock Appl. Phys. Lett. \textbf{79}, 3920 (2001).

\bibitem {2003ApPhL..83..207C}H.~{Callebaut}, S.~{Kumar}, B.~S. {Williams},
Q.~{Hu}, and J.~L. {Reno}, \newblock Appl. Phys. Lett. \textbf{83}, 207 (2003).

\bibitem {2005JAP....97d3702B}O.~{Bonno}, J.-L. {Thobel}, and F.~{Dessenne},
\newblock J. Appl. Phys. \textbf{97}, 043702 (2005).

\bibitem {2007JAP...101h6109J}C.~{Jirauschek}, G.~{Scarpa}, P.~{Lugli}, M.~S.
{Vitiello}, and G.~{Scamarcio}, \newblock J. Appl. Phys. \textbf{101}, 086109 (2007).

\bibitem {2008JAP...104d3101L}H.~{Li}, J.~C. {Cao}, Y.~J. {Han}, X.~G. {Guo},
Z.~Y. {Tan}, J.~T. {L{\"u}},  H.~{Luo}, S.~R. {Laframboise}, and H.~C. {Liu},
\newblock J. Appl. Phys. \textbf{104}, 043101 (2008).

\bibitem {2010ApPhL..96t1110M}A.~{M{\'a}ty{\'a}s}, M.~A. {Belkin}, P.~{Lugli},
and C.~{Jirauschek}, \newblock Appl. Phys. Lett. \textbf{96}, 201110 (2010).

\bibitem {2010ApPhL..96a1103J}C.~{Jirauschek}, \newblock Appl. Phys. Lett.
\textbf{96}, 011103 (2010).

\bibitem {Matyas2009}A.~{M{\'a}ty{\'a}s}, T.~{Kubis}, P.~{Lugli}, and
C.~{Jirauschek}, \newblock Physica E \textbf{42}, 2628 (2010).

\bibitem {2009JAP...105l3102J}C.~{Jirauschek} and P.~{Lugli}, \newblock J.
Appl. Phys. \textbf{105}, 123102 (2009).

\bibitem {2010JAP...107a3104J}C.~{Jirauschek}, A.~{M\'aty\'as}, and
P.~{Lugli}, \newblock J. Appl. Phys. \textbf{107}, 013104 (2010).

\bibitem {2003JAP....93.1586U}T.~{Unuma}, M.~{Yoshita}, T.~{Noda},
H.~{Sakaki}, and H.~{Akiyama}, \newblock J. Appl. Phys. \textbf{93}, 1586 (2003).

\bibitem {1998JAP....84.2112R}P.~{Ramvall}, N.~{Carlsson}, P.~{Omling},
L.~{Samuelson}, W.~{Seifert},  Q.~{Wang}, K.~{Ishibashi}, and Y.~{Aoyagi},
\newblock J. Appl. Phys. \textbf{84}, 2112 (1998).

\bibitem {1999JAP....85.7276A}A.~M. {Alcalde} and G.~{Weber}, \newblock J.
Appl. Phys. \textbf{85}, 7276 (1999).

\bibitem {2005ApPhL..87g2104D}O.~{Drachenko}, J.~{Galibert}, J.~{L{\'e}otin},
J.~W. {Tomm}, M.~P. {Semtsiv},  M.~{Ziegler}, S.~{Dressler}, U.~{M{\"u}ller},
and W.~T. {Masselink}, \newblock Appl. Phys. Lett. \textbf{87}, 072104 (2005).

\bibitem {1993PhRvB..48.8102S}M.~{Sugawara}, N.~{Okazaki}, T.~{Fujii}, and
S.~{Yamazaki}, \newblock Phys. Rev. B \textbf{48}, 8102 (1993).

\bibitem {PhysRevB.40.7714}U.~Ekenberg, \newblock Phys. Rev. B \textbf{40},
7714 (1989).

\bibitem {1987PhRvB..35.7770N}D.~F. {Nelson}, R.~C. {Miller}, and D.~A.
{Kleinman}, \newblock Phys. Rev. B \textbf{35}, 7770 (1987).

\bibitem {2001JAP....89.5815V}I.~{Vurgaftman}, J.~R. {Meyer}, and L.~R.
{Ram-Mohan}, \newblock J. Appl. Phys. \textbf{89}, 5815 (2001).

\bibitem {1993PhRvB..48.2328H}G.~{Hendorfer}, M.~{Seto}, H.~{Ruckser},
W.~{Jantsch}, M.~{Helm},  G.~{Brunthaler}, W.~{Jost}, H.~{Obloh},
K.~{K{\"o}hler}, and D.~J. {As}, \newblock Phys. Rev. B \textbf{48}, 2328 (1993).

\bibitem {2009IJQE...45..1059J}C.~{Jirauschek}, \newblock IEEE J. Quantum
Elect. \textbf{45}, 1059 (2009).

\bibitem {2008PSSCR...5..232K}T.~{Kubis}, C.~{Yeh}, and P.~{Vogl}, \newblock
Phys. Status Solidi C \textbf{5}, 232 (2008).

\bibitem {2005ApPhL..86f2113T}S.~{Tsujino}, A.~{Borak}, E.~{M{\"u}ller},
M.~{Scheinert}, C.~V. {Falub},  H.~{Sigg}, D.~{Gr{\"u}tzmacher},
M.~{Giovannini}, and J.~{Faist}, \newblock Appl. Phys. Lett. \textbf{86},
062113 (2005).

\bibitem {2008ApPhL..93n1103W}A.~{Wittmann}, Y.~{Bonetti}, J.~{Faist},
E.~{Gini}, and M.~{Giovannini}, \newblock Appl. Phys. Lett. \textbf{93},
141103 (2008).

\bibitem {2006ApPhL..89q2120V}A.~{Vasanelli}, A.~{Leuliet}, C.~{Sirtori},
A.~{Wade}, G.~{Fedorov},  D.~{Smirnov}, G.~{Bastard}, B.~{Vinter},
M.~{Giovannini}, and J.~{Faist}, \newblock Appl. Phys. Lett. \textbf{89},
172120 (2006).

\bibitem {oexprJirau01}C.~{Jirauschek}, \newblock Opt. Express \textbf{18},
25922 (2010).

\bibitem {5069088}M.~Razeghi, \newblock IEEE J. Sel. Top. Quant. \textbf{15},
941 (2009).

\bibitem {2008IJQE...44...12Y}M.~{Yamanishi}, T.~{Edamura}, K.~{Fujita},
N.~{Akikusa}, and H.~{Kan}, \newblock IEEE J. Quantum Electron. \textbf{44},
12 (2008).

\bibitem {2010ApPhL..97h1105W}A.~{Wacker}, \newblock Appl. Phys. Lett.
\textbf{97}, 081105 (2010).

\bibitem {2011JAP...109g3112W}M.~{Wienold}, L.~{Schrottke}, M.~{Giehler},
R.~{Hey}, and H.~T. {Grahn}, \newblock J. Appl. Phys. \textbf{109}, 073112 (2011).

\bibitem {2008pssc..5..221J}C.~{Jirauschek} and P.~{Lugli}, \newblock Phys.
Status Solidi C \textbf{5}, 221 (2008).

\bibitem {2007ApPhL..91p1111V}M.~S. {Vitiello}, T.~{Gresch}, A.~{Lops},
V.~{Spagnolo}, G.~{Scamarcio},  N.~{Hoyler}, M.~{Giovannini}, and J.~{Faist},
\newblock Appl. Phys. Lett. \textbf{91}, 161111 (2007).
\end{thebibliography}
\end{document}